\documentclass[10pt,aps,pre,a4paper,twocolumn,superscriptaddress,showpacs,byrevtex,bibnotes]{revtex4-1}
\usepackage{microtype}
\usepackage{color}
\usepackage{amsmath}
\usepackage{amssymb}
\usepackage{braket}
\usepackage{amsthm}
\usepackage{dsfont}
\usepackage{graphicx}
\usepackage[colorlinks = true, citecolor = blue, linkcolor = blue, urlcolor=blue]{hyperref}

\definecolor{green}{rgb}{0.0, 0.5, 0.0}

\definecolor{blue}{rgb}{0.0, 0.0, 0.5}

\newcommand{\cH}{\mathcal{H}}
\newcommand{\cL}{\mathcal{L}}

\begin{document}
\title{Spectral properties of simple classical and quantum reset processes}
	
\author{Dominic C. Rose}
\email[]{dominic.rose@nottingham.ac.uk}
\affiliation{School of Physics and Astronomy, University of Nottingham, Nottingham, NG7 2RD, United Kingdom}
\affiliation{Centre for the Mathematics and Theoretical Physics of Quantum Non-Equilibrium Systems, University of Nottingham, Nottingham, NG7 2RD, United Kingdom}
	
\author{Hugo Touchette}
\email[]{htouchette@sun.ac.za}
\affiliation{National Institute for Theoretical Physics (NITheP), Stellenbosch 7600, South Africa}
\affiliation{Institute of Theoretical Physics, Department of Physics, University of Stellenbosch, Stellenbosch 7600, South Africa}
	
\author{Igor Lesanovsky}
\email[]{igor.lesanovsky@nottingham.ac.uk}
\affiliation{School of Physics and Astronomy, University of Nottingham, Nottingham, NG7 2RD, United Kingdom}
\affiliation{Centre for the Mathematics and Theoretical Physics of Quantum Non-Equilibrium Systems, University of Nottingham, Nottingham, NG7 2RD, United Kingdom}
	
\author{Juan P. Garrahan}
\email[]{juan.garrahan@nottingham.ac.uk}
\affiliation{School of Physics and Astronomy, University of Nottingham, Nottingham, NG7 2RD, United Kingdom}
\affiliation{Centre for the Mathematics and Theoretical Physics of Quantum Non-Equilibrium Systems, University of Nottingham, Nottingham, NG7 2RD, United Kingdom}
	
\begin{abstract}
We study the spectral properties of classical and quantum Markovian processes that are reset at random times to a specific configuration or state with a reset rate that is independent of the current state of the system. We demonstrate that this simple reset dynamics causes a uniform shift in the eigenvalues of the Markov generator, excluding the zero mode corresponding to the stationary state, which has the effect of accelerating or even inducing relaxation to a stationary state. Based on this result, we provide expressions for the stationary state and probability current of the reset process in terms of weighted sums over dynamical modes of the reset-free process. We also discuss the effect of resets on processes that display metastability. We illustrate our results with two classical stochastic processes, the totally asymmetric random walk and the one-dimensional Brownian motion, as well as two quantum models: a particle coherently hopping on a chain and the dissipative transverse field Ising model, known to exhibit metastability.
\end{abstract}
	
\date{\today}
	
\maketitle

\section{Introduction}

The dynamics of stochastic processes, such as animals foraging for food in the wilderness or a person searching for car keys, often include random resets in time, taking the form of returns to past locations where food was successfully located or the last place a person remembers seeing their keys \cite{benichou2011}. Recently, there has been a renewed interest in these processes, due to the fact that they can improve the efficiency of certain random search processes and algorithms in terms of mean hitting or first-passage time \cite{evans2011,evans2011b,evans2013,evans2014,reuveni2016,pal2017}. Reset processes have also been studied from a more physical point of view, as they provide a simple model of nonequilibrium processes breaking detailed balance \cite{manrubia1999,gupta2014,roldan2016,eule2016}, as well as of processes showing dynamical phase transitions in their relaxation dynamics \cite{majumdar2015b}, mean first-passage time \cite{kusmierz2014}, or large deviations \cite{meylahn2015b,Harris2017,hollander2018}. 
				
These studies follow many previous works in mathematics, in queuing theory and in population dynamics, in particular, on stochastic processes involving some form of random resets, variously referred to as failures, catastrophes, disasters or decimations; see, e.g., \cite{brockwell1982,luby1993,pakes1997,kumar2000,cairns2004,janson2012,avrachenkov2013,dharmaraja2015,montero2017}. Most of these works, as well as those from physics mentioned above, make use of the correspondence that exists between resets and renewals to obtain renewal representations of both time-dependent and stationary distributions, in addition to first-passage statistics. Modified Fokker--Planck and Feynman--Kac equations with additional source and sink terms describing the evolution of these distributions and statistics have also been obtained (see, e.g., \cite{evans2011b,dharmaraja2015,meylahn2015b}) and can be solved explicitly for some simple models, including reset versions of Brownian motion \cite{evans2011} and the Ornstein-Uhlenbeck process \cite{pal2015}.
		
In this paper, we present a different approach to reset processes based on the spectral properties of their generator or master operator. Our main result is a relation between the spectrum of the generator of a reset process and that of its reset-free counterpart. More precisely, we show that the real part of the eigenvalues of the generator are shifted for all non-stationary modes by the reset rate, while the corresponding eigenstates, representing the dynamical modes, are not modified. We also provide explicit expressions for the stationary state and current of reset processes involving weighted sums over the dynamical modes, which are applied to two prototypical models, namely, the totally asymmetric random walk in one dimension, related to queuing, and the one-dimensional Brownian motion. The results obtained clearly explain how resets can accelerate or even induce relaxation to a stationary state by opening a spectral gap, and how non-zero stationary currents can be created without having complex eigenvalues in the spectrum of the generator. The eigenvalue result can also be used within the spectral theory of metastability \cite{GaveauSchulman1987Meta,Gaveau1998,Gaveau2006} to demonstrate that weak resetting can modify the weighting of metastable states without modifying those states as such.
		
While resets have been extensively considered in classical nonequilibrium physics, a relatively unexplored area is the addition of resets to quantum systems, either closed or interacting with an environment. This case has been considered recently for a quantum walker subjected to continuous measurements on a particular site, resulting in a random collapse of the wavefunction followed by an evolution starting from the measured site \cite{Thiel2017,Friedman2017}. We conclude our study by generalising our spectral results to this type of open quantum systems described in general by the Lindblad master equation, providing a natural link with the recent extension of the spectral theory of metastability to quantum systems \cite{Macieszczak2016}. We illustrate this generalization by computing numerically the stationary state of a model of coherent hopping in one dimension realizing the reset quantum random walker, and by applying our method to a dissipative transverse field Ising model \cite{Ates2012}, known to display metastability \cite{Rose2016}. 

\section{Reset Markov processes}
\label{ResetMarkov}

We consider a classical stochastic process evolving according to a continuous-time Markov chain. The master equation describing the evolution of the probability $P(C,t)$ for the process to be in state $C$ at time $t$ is given by
\begin{equation}
\partial_t P(C,t)=\sum_{C'\neq C}W(C'\rightarrow C)P(C',t)-R(C)P(C,t),
\end{equation}
where $W(C'\rightarrow C)$ is the transition rate from $C'$ to $C$ and
\begin{equation}
R(C)=\sum_{C'\neq C}W(C\rightarrow C')
\end{equation} 
is the escape rate from $C$. Following the notation commonly used in physics \cite{schutz2001}, this can be written more compactly as
\begin{equation}
\label{MasterEq}
\partial_t \ket{P(t)} = \cL\ket{P(t)},
\end{equation}
where 
\begin{equation}
\ket{P(t)}=\sum_C P(C,t)\ket{C}
\end{equation}
is the probability vector expressed in terms of ket states $\ket{C}$, such that $\left\langle C|C' \right\rangle = \delta_{CC'}$, and
\begin{equation}
\cL=\sum_{C,C'\neq C}W(C\rightarrow C')\ket{C'}\bra{C}-\sum_C R(C)\ket{C}\bra{C}
\end{equation}
is the master operator. 

Since this operator is non-Hermitian, it has two sets of eigenvectors, right and left, given by
\begin{equation}
\cL\ket{r_i}=\lambda_i\ket{r_i}
\end{equation} 
and
\begin{equation}
\bra{l_i}\cL=\lambda_i\bra{l_i},
\end{equation} respectively. These two sets of eigenvectors form a complete basis, are dual to each other, and can be normalised in a such a way that $\left\langle l_i | r_j \right\rangle=\delta_{ij}$. 

We assume here that the process is ergodic and, therefore, that it has a unique stationary state $\ket{P_\text{ss}}$, corresponding from \eqref{MasterEq} to the eigenvalue $\lambda_1=0$, so that $\ket{P_\text{ss}}=\ket{r_1}$. We also assume that $\cL$ contains no Jordan blocks, corresponding to non-exponentially decaying modes, so we do not need to consider generalized eigenvectors \footnote{The left eigenvectors may not be calculable in the case where there are Jordan blocks. The right eigenvectors and their corresponding eigenvalues are however the same and the stationary state can still be constructed using an alternative approach based on the renewal representation of trajectory ensembles.}. The normalization of the stationary state can be expressed as
\begin{equation}
\sum_C \left\langle C|P_\text{ss}\right\rangle=\left\langle-|P_\text{ss}\right\rangle=1,
\end{equation} 
where we have introduced the ``flat'' state
\begin{equation}
\bra{-}=\sum_C \bra{C}.
\end{equation} 
Conservation of probability also requires $\bra{-}\cL=0$, which implies $\bra{l_1}=\bra{-}$ and hence $\left\langle -|r_i \right\rangle=0$ for all $ i \neq 1$. From the ergodicity assumption, all other eigenvalues are possibly complex but have real parts less than zero, that is, $\text{Re}(\lambda_i)<0$ for all $i\neq1$. Both this and the Jordan block assumption can be relaxed to arrive at similar but slightly more general results.
		
The generator $\cL$ defines our original process. The reset version of that process is constructed simply by adding new transitions at a rate $\Gamma$ from every configuration to a target or reset state, denoted by $C_0$. The generator of the reset process is thus given by
\begin{eqnarray}
\cL^\Gamma=\cL+\Gamma\sum_{C\neq C_0} \ket{C_0}\bra{C}-\Gamma\sum_{C\neq C_0}\ket{C}\bra{C},
\end{eqnarray}
where the additional terms can be absorbed into the old jump operators and escape rate operator to give shifted transition and escape rates. Note that we can add 
\begin{equation}
0=\Gamma\ket{C_0}\bra{C_0}-\Gamma\ket{C_0}\bra{C_0}
\end{equation} 
to $\cL^\Gamma$ to obtain the simpler form
\begin{align}
\cL^\Gamma&=\cL+\Gamma\sum_{C} \ket{C_0}\bra{C}-\Gamma\sum_{C}\ket{C}\bra{C}\nonumber\\
&=\cL+\Gamma\ket{C_0}\bra{-}-\Gamma I,
\label{LR}
\end{align}
where $I$ is the identity operator. In this form, it is clear that the reset adds transitions from all states to $C_0$, contributing to an extra escape rate $\Gamma$ in the diagonal, which keeps the conservation condition $\bra{-}\cL^\Gamma=0$.

The dynamics generated by \eqref{LR} is arguably the simplest form of reset process. As defined, reset events are Poissonian with rate $\Gamma$, which is independent of the state of the system at the time of the reset. 

\section{Results}
\label{SecRes}

We study in this section the spectral properties of the reset process, deriving the new eigenvalues, left eigenvectors and right eigenvectors of $\cL^\Gamma$ in terms of those of $\cL$. The results are then used to obtain spectral representations of the stationary state and current of the reset process, and to discuss the effect of resets on metastable states.
		
\subsection{Eigenvalues and eigenvectors}
\label{ResetSpec}

We begin our analysis with the right eigenvectors by noting that, due to conservation of probability and the fact that $\left\langle -|r_i \right\rangle=0$ for all $i\neq1$, we have 
\begin{align}
\cL^\Gamma\ket{r_i} &=\big(\cL+\Gamma\ket{C_0}\bra{-}-\Gamma I\big)\ket{r_i}\nonumber\\
&=(\lambda_i-\Gamma)\ket{r_i}
\end{align}
for $i\neq 1$. Consequently, the right eigenvectors of the reset process are the same as those of the original process, while the eigenvalues are shifted down by $\Gamma$:
\begin{equation}
\lambda^\Gamma_i = \lambda_i-\Gamma,\qquad i\neq 1.
\end{equation} 
This applies, as noted, to all modes except the stationary state, discussed below, which is still such that $\lambda^\Gamma_1=0$.
			
To determine the left eigenvectors, we act from the left with the original eigenmodes
\begin{align}
\bra{l_i}\cL^{\Gamma}&=\bra{l_i}\big(\cL+\Gamma\ket{C_0}\bra{-}-\Gamma I\big)\nonumber\\
&=(\lambda_i-\Gamma)\bra{l_i}+\Gamma\left\langle l_i|C_0\right\rangle\bra{-}\nonumber\\
&=(\lambda_i-\Gamma)\left(\bra{l_i}+\frac{\Gamma\left\langle l_i|C_0\right\rangle}{\lambda_i-\Gamma}\bra{-}\right).\nonumber
\end{align}
But by conservation of probability $\bra{-}\mathcal{L}^\Gamma=0$, we also have
\begin{equation}
\bra{l_i}\cL^{\Gamma}=\left(\bra{l_i}+\frac{\Gamma\left\langle l_i|C_0\right\rangle}{\lambda_i-\Gamma}\bra{-}\right)\cL^{\Gamma}
\end{equation}
for all $i\neq 1$. Thus we see that the new left eigenvectors are given by
\begin{eqnarray}
\label{NewLeft}
\bra{l^\Gamma_i}=\bra{l_i}+\frac{\Gamma\left\langle l_i|C_0\right\rangle}{\lambda_i-\Gamma}\bra{-},\qquad i\neq1.
\end{eqnarray}
For $i=1$, we have as before $\bra{l_1^\Gamma}=\bra{-}$.

\subsection{Stationary state}
\label{StatStateResets}

The stationary state $\ket{P^\Gamma_\text{ss}}$ of the reset process, corresponding to $\ket{r_1^\Gamma}$, is obtained from the results above by noting that the new left and right eigenvectors are orthonormal to each other, so that $\left\langle{l^\Gamma_i}|P^\Gamma_\text{ss}\right\rangle=\delta_{i1}$. Substituting this condition in Eq.~\eqref{NewLeft}, we find
\begin{equation}
\left\langle{l_i}|P^\Gamma_\text{ss}\right\rangle=-\frac{\Gamma\left\langle l_i|C_0\right\rangle}{\lambda_i-\Gamma},
\end{equation}
for $i\neq1$, and thus
\begin{equation}
\label{NewSteadyState}
\ket{P^\Gamma_\text{ss}}=\ket{P_\text{ss}}-\sum_{i=2}^{D}\frac{\Gamma\left\langle l_i|C_0\right\rangle}{\lambda_i-\Gamma}\ket{r_i}.
\end{equation}
	
We see that in addition to the stationary state of the process without resets, the resetting stationary state contains a contribution of the dynamical modes, weighted according to how significant they are in the evolution of the reset state in the original dynamics (the overlaps $\left\langle l_i|C_0\right\rangle$), and the reset rates magnitude relative to the corresponding eigenvalue $\lambda_i$. This agrees with intuition: since the eigenvalues are related to the lifetimes of the dynamical modes, if the average time between resets is larger than the lifetime of a mode, it will not make a significant contribution to the new steady state. We note that this equation also has the expected limits of $\ket{P^0_\text{ss}}=\ket{P_\text{ss}}$ and $\ket{P^\infty_\text{ss}}=\ket{C_0}$.
		
This result applies for processes with a finite number $D$ of states, but also to infinite-dimensional processes, provided that they possess a well-defined spectrum of eigenvalues with corresponding left and right eigenmodes. This latter fact will be illustrated in the next section with the example of Brownian motion. 

We should also note that the result holds if the reset state $\ket{C_0}$ is replaced by a reset probability distribution $\ket{P_0}$ over configurations, giving the probability of reaching different states after a reset event. In this case, $\ket{C_0}$ in Eq.~\eqref{NewSteadyState} is simply replaced by the ``mixed'' reset state $\ket{P_0}$. This follows since the reset state is never referred to above as anything more than as a vector in state space.

Finally we mention the modification resulting from a lack of ergodicity, i.e., when $\mathcal{L}\ket{r_i}=0$ and $\bra{l_i}\mathcal{L}=0$ for more than one pair of states, causing an initial state dependence of the stationary state. Generically, we can still choose a basis such that $\bra{l_1}=\bra{-}$ and $\left\langle l_i|r_j\right\rangle=\delta_{ij}$ within this null eigenspace, allowing us to use nearly the same proof scheme as above. The reset rates break the zero eigenvalues degeneracy and provides a unique stationary state, with the remainder of the null eigenspace shifted to an eigenvalue of $-\Gamma$. In Eq. \eqref{NewSteadyState} the unique steady state $\ket{P_{\text{ss}}}$ is replaced by the state-dependent steady state that would be reached from the reset state under the original dynamics.
\subsection{Stationary current}
\label{DetailedBalance}
			
The current associated with the stationary state $\ket{P^\Gamma_\text{ss}}$ of the reset process is defined, for any given link or transition $c\rightarrow c'$, by
\begin{equation}
J^\Gamma_{c\rightarrow c'}=\left\langle c|P^\Gamma_{\text{ss}}\right\rangle\cL^\Gamma_{c'c}-\left\langle c'|P^\Gamma_{\text{ss}}\right\rangle\cL^\Gamma_{cc'},
\end{equation} 
where $\cL^\Gamma_{ij}$ is the $(i,j)$ component of $\cL^\Gamma$. Substituting the expression of the stationary state, found in Eq.~\eqref{NewSteadyState}, together with the expression of the generator $\cL^\Gamma$, we can decompose the current into three parts as
\begin{align}
J^\Gamma_{c\rightarrow c'}&=J_{c\rightarrow c'}
	+\sum_{i=2}^{D}\frac{\Gamma\left\langle l_i|C_0\right\rangle}{\Gamma-\lambda_i}J^{i}_{c\rightarrow c'}\nonumber\\
	&\quad+\Gamma\left(\left\langle c|P^\Gamma_{\text{ss}}\right\rangle{\delta}_{c'c_0}-\left\langle c'|P^\Gamma_{\text{ss}}\right\rangle{\delta}_{c_0c'}\right),
\label{eqcurr1}
\end{align}
where
\begin{equation}
J^{i}_{c\rightarrow c'}=\left\langle c|r_i\right\rangle\cL_{c'c}-\left\langle c'|r_i\right\rangle\cL_{cc'}.
\end{equation}
The first term on the right-hand side of Eq.~\eqref{eqcurr1} is the current of the original reset-free process, while the second is the weighted contribution of new currents coming from the non-stationary modes of the original process. Finally, the third term is the current coming from the reset transitions. From the signs appearing in the last term, we see that there is a current loop from all the states to the reset state $C_0$ and then back from $C_0$ to all other states, so that $C_0$ acts both as a sink and source.
			
From this result, it is clear that the reset process will violate the condition of detailed balance, i.e., $J^\Gamma\neq 0$, if the original process satisfies the condition of detailed balance, i.e., $J=0$. In this case, it is known that the original process has a real spectrum, which implies from our results that the spectrum of the reset process must also be real, even though $J^\Gamma\neq 0$. This shows that currents are not necessarily associated with complex spectra of the dynamical generator. In fact, detailed balance is only a sufficient condition for the generator spectrum to be real, not a necessary condition.

In principle, it is also possible to have a reset process satisfying detailed balance ($J^\Gamma=0$) if the original process violates detailed balance ($J\neq 0$). However, this is a rather peculiar case, requiring that the added reset transitions with rate $\Gamma$ exactly counterbalance all the non-zero currents arising in the reset-free process. In this case, the original process must again have a real spectrum in order for the spectrum of the reset process to be real.
		
\subsection{Metastability}
\label{Metastability}
		
We close this section by discussing the effect of resets on metastable states that arise when the evolution of the probability distribution $P(C,t)$ exhibits two distinct time scales: a fast evolution towards long-lived metastable states, followed by a slow relaxation to the final stationary state. These metastable states typically reside in a reduced subset of the full state space, called the \textit{metastable manifold} (MM), with the later relaxation to the stationary state occurring within the MM (see, e.g., \cite{Macieszczak2016} and references therein for definitions and nomenclature that applies both to classical and quantum metastability).
			
Much work has been done on Markovian processes to understand metastability \cite{GaveauSchulman1987Meta,Gaveau1998,Bovier2002,Gaveau2006,larralde2005,olivieri2005}, based on the presence of large gaps in the spectrum of the master operator, which are necessary for the occurrence of distinct timescales. The MM in this context is understood to correspond to the reduced set of eigenmodes defined by these gaps, with the late time relaxation given by a projection of the master operator onto the MM.

To illustrate this phenomenon in the simplest way possible, let us consider a Markov process with a unique stationary state and a large gap between the second and third eigenvalues, i.e., $|\text{Re}(\lambda_2)|\ll|\text{Re}(\lambda_3)|$. The MM of long-lived states corresponds in this case to a one-dimensional manifold of linear combinations of the stationary state and $\ket{r_2}$, with the coefficients of $\ket{r_2}$ bounded by the maximum and minimum values of $\bra{{l_2}}$ on the space of probability distributions, i.e., the maximum and minimum components of this vector in the configuration basis $c_2^{\text{max}}$ and $c_2^{\text{min}}$. These two values define the so-called \textit{extreme metastable states} (eMSs) on the boundary of the manifold:
\begin{align}
\ket{\tilde{P}_1}&=\ket{P_\text{ss}}+c_2^{\text{max}}\ket{r_2}\nonumber\\
\ket{\tilde{P}_2}&=\ket{P_\text{ss}}+c_2^{\text{min}}\ket{r_2},
\label{ems}
\end{align}
in terms of which we can write the stationary state as
\begin{align}
\ket{P_\text{ss}}&=\frac{1}{\Delta c_2}\left(-{c}_{2}^{\rm min}\ket{\tilde{P}_1}+{c}_{2}^{\rm max}\ket{\tilde{P}_2}\right)\label{SteadyStateMMExpansion}\\
	&=p_1^\text{ss}\ket{\tilde{P}_1}+p_2^\text{ss}\ket{\tilde{P}_2},\nonumber
\end{align}
where $\Delta c_2={c}_{2}^{\rm max}-{c}_{2}^{\rm min}$. Note that $c_2^\text{min}\leq0$ as $\bra{l_2}$ is orthogonal to the stationary state which has purely positive components, so that the coefficients $p_1^\text{ss}$ and $p_2^\text{ss}$ in this expansion can be viewed as the probability weight of the two eMSs. Finally, we can construct an effective evolution on this subspace by projecting the master operator to find an effective master operator given in terms of $\lambda_2$ and the maximum and minimum components of $\bra{l_2}$
\begin{equation}
\cL_{\text{eff}}=\frac{-\lambda_{2}}{\Delta{c}_{2}}
\begin{pmatrix}
-{c}_{2}^{\rm max} & -{c}_{2}^{\rm min}\\
{c}_{2}^{\rm max} & {c}_{2}^{\rm min}
\end{pmatrix}.
\label{EffectiveMMDynamics}
\end{equation}

The meaning of the above is the following. Consider the system starting in an initial state $\ket{P_0}$ (either a specific configuration or a probability over configurations). At some time $t$, the state of the system will read in terms of the spectrum of ${\mathcal L}$,
\begin{equation}
\ket{P_0} = \ket{P_\text{ss}} + e^{t \lambda_2} \langle l_2 | P_0 \rangle \ket{r_2} + \sum_{i \geq 3} 
e^{t \lambda_i} \langle l_i | P_0 \rangle \ket{r_i} .
\label{pot}
\end{equation}
Due to the separation of timescales, for times $t$ such that 
$ 1/|\text{Re}(\lambda_3)|\ll t \ll 1/|\text{Re}(\lambda_2)|$, all but the first two terms in \eqref{pot} will be negligible, assuming that the overlap of the initial state with the modes $i\geq 3$ is small so that these terms are suppressed by the decaying modes for $i\geq 3$. Within these timescales, the initial state $\ket{P_0}$ evolves to a state in the one-dimensional MM, subsequently evolving within the MM and eventually reaching the unique stationary state for $t \gg 1/|\text{Re}(\lambda_2)|$, i.e., schematically
\begin{equation}
\ket{P_0} \to p_1(t) \ket{\tilde{P}_1} + p_2(t) \ket{\tilde{P}_2} \to \ket{P_{\rm ss}} . 
\end{equation}
The evolution for $t \gg 1/|\text{Re}(\lambda_3)|$, prior to reaching the stationary state, is within the one-dimensional MM, as it corresponds to the evolution of $p_{1,2}(t) \geq 0$ in the linear combination above, with $p_1(t)+p_2(t)=1$, and is described by the effective generator \eqref{EffectiveMMDynamics}.
			
We now add resets to this metastable dynamics, focusing for simplicity on the one-dimensional metastable manifold case. First, we note that we can rewrite Eq.~\eqref{NewSteadyState} as
\begin{equation}
\ket{P^\Gamma_\text{ss}}=\ket{P_\text{ss}}-\sum_{i=2}^{D}\frac{\left\langle l_i|C_0\right\rangle}{\frac{\lambda_i}{\Gamma}-1}\ket{r_i}.
\end{equation}
As a result, we see that, if we consider $\Gamma \approx |\lambda_2|\ll|\lambda_3|$, then the coefficients in the sum for the terms $i\geq3$ are small compared to the coefficient for $i=2$, so we can truncate to only the first two terms:
\begin{equation}\label{MetaTruncation}
\ket{P^\Gamma_\text{ss}}\approx\ket{P_\text{ss}}-\frac{\Gamma\left\langle l_2|C_0\right\rangle}{\lambda_2-\Gamma}\ket{r_2}.
\end{equation}
Since the spectrum of the process with resets is simply a real shift by $\Gamma$, values on this scale preserve the gap in the spectrum required for metastability, as for $\Gamma\approx|\lambda_2|$ we still have $\text{Re}(\lambda_2)-\Gamma\ll \text{Re}(\lambda_3)-\Gamma$. Physically, this regime corresponds to the average time between resets being comparable to the lifetime of the metastable phases, in which the short timescale dynamical modes represented by $i\geq3$ are averaged out and make negligible contribution to the stationary state. 

Using the definitions \eqref{ems} of the eMSs with the modified left eigenmodes from Sec. \ref{ResetSpec}, the eMSs of the model with resets are given by
\begin{align}\label{MetaeMS}
\ket{\tilde{P}^\Gamma_{1}}=&\ket{{P}^\Gamma_{\rm ss}}+{c}_{2}^{\Gamma,\rm max}\ket{{r}_{2}}\nonumber\\
\ket{\tilde{P}^\Gamma_{2}}=&\ket{{P}^\Gamma_{\rm ss}}+{c}_{2}^{\Gamma,\rm min}\ket{{r}_{2}},
\end{align}
where, since the modification to the left eigenmodes is simply a shift by the flat state, the coefficients are now given by
\begin{equation}
{c}_{2}^{\Gamma,\rm min/max}={c}_{2}^{\rm min/max}-\frac{\Gamma\left\langle l_2|C_0\right\rangle}{\Gamma-\lambda_2}.
\end{equation}
Substituting this and Eq.~\eqref{MetaTruncation} into Eq.~\eqref{MetaeMS}, we see that the eMSs with reset are approximately equal to the original eMSs, as the effect of the modifications to the steady state and left eigenmodes cancel. Applying Eq.~\eqref{SteadyStateMMExpansion} to the model with reset and using $\ket{\tilde{P}^\Gamma_{i}}\approx\ket{\tilde{P}_{i}}$, we thus obtain
\begin{eqnarray}\label{ResetMetastableDecomp}
\ket{P^\Gamma_\text{ss}}&\approx&\left(p_1^\text{ss}+\Delta p\right)\ket{\tilde{P}_1}+\left(p_2^\text{ss}-\Delta p\right)\ket{\tilde{P}_2}\nonumber\\
&=&\ket{P_{ss}}+\Delta p \ket{\tilde{P}_1}-\Delta p\ket{\tilde{P}_2},\nonumber
\end{eqnarray}
where
\begin{equation}
\Delta p = \frac{\Gamma\left\langle l_2|C_0\right\rangle}{\Delta c_2(\Gamma-\lambda_2)}.
\end{equation}
We see that, depending on the overlap of the reset state with $\ket{l_2}$, this coefficient can cause a notable modification of the steady state mixture even for small $\Gamma$. This means physically that resets will make whichever eMS is closer to the reset state more likely to occur in the stationary state, as expected. Finally, the modified effective dynamics can be constructed simply by replacing the coefficients and eigenvalues with the new ones in Eq.~\eqref{EffectiveMMDynamics}.

\section{Applications}
\label{Applications}

We apply in this section our formula \eqref{NewSteadyState} for the stationary state of the reset process for two exactly-solvable models. The applicability of this formula is obviously limited by the fact that it requires the full spectrum of the reset-free process. For this reason, we expect it to be more useful for approximating the stationary state than for calculating that state exactly, either by truncating the sum involved to a limited number of modes or by expanding the sum perturbatively in $\Gamma$. Moreover, while exact results can be hard to find, the formula can be useful numerically when applied to processes in which the resets break symmetries of the original, reset-free process. Such symmetries can indeed be used to diagonalise the original process for system sizes much larger than would otherwise be possible, with the resulting spectrum then being used in \eqref{NewSteadyState} to derive the stationary state with resets. This is demonstrated later in the context of adding resets to a closed quantum system breaking time-reversal symmetry.
		
\subsection{Totally asymmetric random walk}
\label{AsymHop}	
			
The first model that we consider is a particle hopping on a one-dimensional lattice of length $L$ with periodic boundary conditions, so the states are $\ket{x}$ with $x\in\{1,2,\ldots,L\}$ and $\ket{L+1}=\ket{1}$. We take the particle to hop only to the right with rate $\gamma$, so that the master operator is
\begin{equation}
\cL=\gamma\sum_{x=1}^{L}\ket{x+1}\bra{x}-\gamma I.
\label{HoppingMastOp}
\end{equation}
This operator is translation invariant and can be diagonalized by discrete Fourier transform, with left and right eigenvectors given by
\begin{equation}
\ket{r_n}=\frac{1}{L}\sum_{x=1}^{L}e^{i\frac{2\pi n}{L}(x-1)}\ket{x}
\end{equation}
and
\begin{equation}
\bra{l_n}=\sum_{x=1}^{L}e^{-i\frac{2\pi n}{L}(x-1)}\bra{x},
\end{equation}
respectively, where $n\in\{0,...,L-1\}$. As before, these eigenvectors are normalized such that $\left\langle l_n | r_m \right\rangle = \delta_{nm}$. Moreover, the eigenvalues are given by 
\begin{equation}
\lambda_n = \gamma(e^{-i2\pi n/L}-1).
\end{equation}

We now add resets at rate $\Gamma$ onto the reset site $\ket{1}$. The new eigenvalues and eigenvectors can then be calculated exactly using the results of the previous section, with Eq.~\eqref{NewSteadyState} leading to 
\begin{equation}
\left\langle x|P_{\text{ss}}^\Gamma\right\rangle = \frac{\Gamma}{\gamma+\Gamma}S_L(x),
\end{equation}
where
\begin{equation}
S_L(x)=\frac{1}{L}\sum_{n=0}^{L-1}\frac{e^{i\frac{2\pi n}{L}x}}{e^{i\frac{2\pi n}{L}}-\frac{\gamma}{\gamma+\Gamma}}.
\end{equation}
Considering the sum $S_L(x)$ on different sites, we find
\begin{equation}
S_L(x+1)-\frac{\gamma}{\gamma+\Gamma}S_L(x)=\delta_{xL},
\end{equation}
and so $S_L(x+1)=\gamma/(\gamma+\Gamma)S_L(x)$ when $x\neq L$. Consequently, 
\begin{equation}
\left\langle x|P_{\text{ss}}^\Gamma\right\rangle = \frac{\Gamma}{\gamma}{\left(\frac{\gamma}{\gamma+\Gamma}\right)}^{x}S_L(1).
\end{equation}
At this point, rather than explicitly calculate the sum $S_L(1)$, we can just normalise the stationary state to find
\begin{equation}
S_L(1)=\frac{1}{1-{\left(\frac{\gamma}{\gamma+\Gamma}\right)}^{L}},
\end{equation}
thus giving
\begin{equation}\label{NewHoppingStationary}
P_\text{ss}^\Gamma(x)=\left\langle x|P_{\text{ss}}^\Gamma\right\rangle = \frac{\Gamma}{\gamma+\Gamma}\frac{1}{1-{\left(\frac{\gamma}{\gamma+\Gamma}\right)}^{L}}{\left(\frac{\gamma}{\gamma+\Gamma}\right)}^{x-1}.
\end{equation}

This result is interesting because, while the infinite-size limit of the initial model neither has a stationary state or the ability to reach one (the spectral gap tends to zero), the infinite-size limit of the model with resets gains a gap of exactly $\Gamma$, with the corresponding stationary state given by the limit of Eq.~\eqref{NewHoppingStationary}, with probabilities
\begin{equation}\label{AsymHopSS}
P_\text{ss}^\Gamma(x) = \frac{\Gamma}{\gamma+\Gamma}{\left(\frac{\gamma}{\gamma+\Gamma}\right)}^{x-1},
\end{equation}
where normalization can be checked via the geometric series. In this way, we see that resets localise the particle near the reset state, with a localisation length of
\begin{equation}
\epsilon = \frac{1}{\ln\left(\frac{\gamma+\Gamma}{\gamma}\right)}.
\end{equation}

\subsection{Brownian motion}
\label{BMReset}

We now show how to apply our results to continuous-state models by seeing them as the limit of a sequence of finite-dimensional models which have a discrete spectrum and well-defined eigenvectors. We consider for this purpose the reset Brownian motion in one dimension, first studied in \cite{evans2011} via a modified Fokker-Planck equation. 

The master, or Fokker-Planck operator in this case, without reset is the Laplacian
\begin{equation}
\cL p(x)=D\frac{d^2 p(x)}{d x^2}
\end{equation}
on the real line, with $D$ as the diffusion constant and $p(x)$ the probability density. The spectrum of this operator is trivial, but it has no normalised stationary state, nor is it possible to define the orthogonality relation between left and right eigenfunctions. To address this problem, we restrict the system, as commonly done in physics, to the finite interval $[-L/2,L/2)$ with periodic boundary conditions. The spectrum of this restricted model is simply given by
\begin{equation}
\lambda_n=-D\left(\frac{2\pi n}{L}\right)^2,\qquad n\in\mathbb{Z},
\end{equation}
with equal right and left eigenmodes, due to the Hermiticity of $\cL$, given by
\begin{equation}
r_n(x)=\frac{1}{\sqrt{L}}e^{i\frac{2\pi n}{L}x},
\end{equation}
and normalised according to
\begin{eqnarray}
\left\langle r_n|r_m \right\rangle = \int_{-L/2}^{L/2}r^*_n(x)r_m(x)\, dx=\delta_{nm}.
\end{eqnarray}

Since the spectrum exists and can be normalized appropriately, we may use our results of Sec.~\ref{SecRes}. Adding resets at a rate $\Gamma$ to the position 0, and defining the states $\ket{0}$ and $\bra{-}$ by $\left\langle g|0 \right\rangle=g(0)$ and 
\begin{equation}
\left\langle -|f \right\rangle=\left\langle l_0|f \right\rangle=\int_{-L/2}^{L/2}f(x)\, dx,
\end{equation}
respectively, we find the new stationary state, which is now a probability density, to be given by the sum
\begin{align}
p_{\text{ss}}^\Gamma(x)=\frac{\Gamma}{L}\sum_{n=-\infty}^{\infty}\frac{e^{i\frac{2\pi n}{L}x}}{D\left(\frac{2\pi n}{L}\right)^2+\Gamma}.
\end{align}
Taking the infinite-size limit, we then find
\begin{equation}
p_{\text{ss}}^\Gamma(x)=\Gamma\int_{-\infty}^{\infty}\frac{dk}{2\pi}\frac{e^{ikx}}{Dk^2+\Gamma},
\end{equation}
which can be solved using residues to give
\begin{equation}
p_{\text{ss}}^\Gamma(x)=\frac{1}{2}\sqrt{\frac{\Gamma}{D}}e^{-\sqrt{\frac{\Gamma}{D}}|x|}.
\end{equation}
This agrees with the result of \cite{evans2011} and is similar to the distribution \eqref{AsymHopSS} found for the discrete random walk.

\section{Open quantum systems with resets}
\label{OQS}

We next consider adding resets to open quantum systems interacting with an environment. Under appropriate conditions on the timescales of the dynamics in the environment and the strength of the interactions, the environment can be suitably viewed as memoryless, allowing us to consider the system to be a quantum generalization of the Markovian systems considered earlier \cite{{Lindblad1976Equation, Gardiner2004QNoise, Breuer2002OpenQuant}}. 
		
\subsection{Theory}
\label{secresqtheory}
		
We consider a quantum system in a Hilbert space $\cH$ of dimension $\text{dim}(\cH)$ with density matrix $\rho$, whose evolution is given by
\begin{equation}
\frac{d\rho}{dt}=\cL(\rho),
\label{eqlindeq}
\end{equation}
where
\begin{equation}\label{LindbladOp}
\cL(\rho) = -i[H,\rho]+\sum_{j}\left[{J}_{j}\rho{J}_{j}^{\dagger}-\frac{1}{2}\{{J}_{j}^{\dagger}{J}_{j},\rho\}\right],
\end{equation}
is the Lindblad master operator. Here $H$ is the Hamiltonian of the system and the jump operators ${J}_{j}$ mediate the system-bath interaction, providing coupling of the system to the surrounding environment. 

Since $\cL$ acts linearly on the density matrix, the evolution \eqref{eqlindeq} can be understood in terms of its eigenvalues and eigenmatrices. Let us denote the eigenvalues of $\cL$ by ${\lambda}_{k}$ and order them such that $\text{Re}({\lambda}_{k})\geq\text{Re}({\lambda}_{k+1})$. As in the classical case, we have $\text{Re}(\lambda_k)\leq 0$, with $\lambda_1=0$ corresponding to the stationary state, due to the fact that $\cL$ is completely positive and trace-preserving. Moreover, as $\cL$ is in general not Hermitian, it has right and left eigenmatrices denoted by 
\begin{equation}
\cL(R_k)=\lambda_k R_k
\end{equation}
and
\begin{equation}
\cL^\dagger(L_k)=\lambda_k^* L_k,
\end{equation} 
respectively. These are normalised such that
\begin{equation}
\text{Tr}({L}^\dagger_{k}{R}_{k'})={\delta}_{kk'}.
\end{equation}
Generally, the stationary state ${\rho}_{\rm ss}$ is unique. Normalising it, as usual, by $\text{Tr}({\rho}_{\rm ss})=1$, we then have $L_1=\mathds{1}$. Defining
\begin{equation}
{c}_{k}=\text{Tr}[{L}^\dagger_{k}\rho(0)]
\end{equation} 
for an initial state $\rho(0)$, the system's state at time $t$ is given by
\begin{equation}\label{Expansion}
\rho(t)={e}^{t\cL}[\rho(0)]={\rho}_{\rm ss}+\sum_{k}{c}_{k}{e}^{t{\lambda}_{k}}{R}_{k}.
\end{equation}

For the classical stochastic processes discussed in the previous sections, reset occurred with equal probability from every state at times distributed exponentially with rate $\Gamma$. We can construct a similar kind of reset dynamics for quantum open systems by adding jump operators
\begin{equation}
J^\Gamma_i=\sqrt{\Gamma}\ket{\psi}\bra{\phi_i}
\end{equation} 
to the Lindbladian \eqref{LindbladOp}. Here $\ket{\phi_i}$ form a complete orthonormal basis, i.e., $\left\langle\phi_i|\phi_j\right\rangle=\delta_{ij}$, and $\ket{\psi}$ is the reset state. This modifies the Lindblad generator to
\begin{equation}\label{ModifiedLindblad}
\cL_{\Gamma}(\rho) = \cL(\rho)+\mathcal{V}_{\Gamma}(\rho)-\Gamma\rho,
\end{equation}
where
\begin{equation}
\mathcal{V}_{\Gamma}(\rho) = \Gamma\,\text{Tr}(\rho)\ket{\psi}\bra{\psi}. 
\end{equation}
To check that this construction has state-independent resets, as desired, we can consider the quantum jump Monte Carlo (QJMC) approach to simulating individual trajectories of the system's evolution (see, e.g., \cite{plenio1998}). When the system undergoes a stochastic dissipative change (a ``jump''), the probability of the change being a reset is
\begin{equation}
P_\Gamma(\ket{\phi})\propto\sum_k^{\text{dim}(\mathcal{H})}\bra{\phi}{J^\Gamma_k}^\dagger J^\Gamma_k\ket{\phi}=\Gamma\left\langle\phi|\phi\right\rangle,
\end{equation}
and is thus state independent as required.

To analyse the spectrum of the model with resets, we make the same assumptions as in Sec. \ref{ResetMarkov}: we assume that the stationary state of the model without resets is unique, and that it is diagonalizable (i.e., there are no non-trivial Jordan blocks). The uniqueness of the stationary state implies $\text{Tr}(R_i)=0$ and hence
\begin{equation}
\cL_\Gamma(R_i)=(\lambda_i-\Gamma)R_i
\end{equation} 
for $i\in\{2,...,\text{dim}({\cH}^2)\}$. This shows, similarly to the classical case, that the $R_i$'s remain eigenmodes of the model with resets, with modified eigenvalues
\begin{equation}
\lambda^\Gamma_i=\lambda_i-\Gamma.
\end{equation} 
For $i=1$, we have again $\lambda_1=0$ and the stationary state $R_1=\rho_{\rm ss}$, which we calculate below using an analogous method to that of Sec. \ref{ResetSpec}.
		
Next, we consider the corresponding left eigenmodes dual to the above. The adjoint equation is given by 
\begin{equation}
\cL_{\Gamma}^\dagger(\rho) = \cL^\dagger(\rho)+\mathcal{V}_{\Gamma}^\dagger(\rho)-\Gamma\rho,
\end{equation}
where $\mathcal{V}_{\Gamma}^\dagger(\rho) = \Gamma\bra{\psi}\rho\ket{\psi}I$. Note that the identity $I$ remains an eigenmode with eigenvalue 0, as expected. Inserting the original left eigenmodes, we find 
\begin{equation}
\cL_\Gamma^\dagger(L_k)=(\lambda^*_k-\Gamma)L_k+\Gamma\bra{\psi}L_k\ket{\psi}{I}.
\end{equation}
Defining
\begin{equation}
L^\Gamma_k=L_k+\frac{\Gamma\bra{\psi}L_k\ket{\psi}}{\lambda^*_k-\Gamma}I,
\end{equation}
we see that, since the identity is annihilated by the adjoint operator, we have
\begin{equation}
\cL_\Gamma^\dagger\left(L^\Gamma_k\right)=(\lambda^*_k-\Gamma)L^\Gamma_k,
\end{equation}
demonstrating that the new left eigenmodes are $L^\Gamma_k$. It can be checked that $\text{Tr}({L^\Gamma_i}^\dagger R_j)=\delta_{ij}$ for $j\neq 1$ and for all $i$, as expected, since $\text{Tr}(R_j)=0$. Finally, requiring $\text{Tr}({L^\Gamma_i}^\dagger \rho^\Gamma_{ss})=\delta_{i1}$ for the new stationary state $\rho^\Gamma_{ss}$ gives the expansion coefficients of that state in the original right eigenmode basis as
\begin{equation}\label{SteadyStateResets}
\rho_{ss}^\Gamma=\rho_{ss}^0-\sum_{j=2}^{\text{dim}(\cH)^2}\frac{\Gamma\bra{\psi}L^\dagger_j\ket{\psi} }{\lambda_j-\Gamma}R_j,
\end{equation}
which, as can be checked, gives $\cL_\Gamma\rho_{ss}^\Gamma=0$. 

The practical applicability of these results depends, as in the classical case, on the system studied and whether, in particular, it has symmetries simplifying the spectral problem. There is an additional benefit in the case of closed quantum systems coming from the fact that the Lindblad equation reduces in that case to the von Neumann equation, allowing for much larger system sizes to be studied compared to a direct spectral solution of the open quantum problem. We demonstrate this next.

\subsection{Coherent hopping on a chain}
\label{CoherentHop}

The first example that we consider to illustrate our results is a simple model of coherent hopping on a closed periodic chain, described by the Hamiltonian 
\begin{align}
H=&\gamma\sum_{x=1}^{L-1} \left( \ket{x+1}\bra{x}+\ket{x}\bra{x+1} \right) 
\nonumber \\
&+ \gamma \left( \ket{1}\bra{L}+\ket{L}\bra{1} \right),
\end{align}
and no jump operators, so that the reset-free system is closed. The dynamics generated by this Hamiltonian is similar to that considered in \cite{{Thiel2017,Friedman2017}}. The main difference is that we consider state-independent resets whereas \cite{{Thiel2017,Friedman2017}} consider resets induced by continuously measuring the system's state on a particular site, leading to a single jump operator proportional to the projective measurement on that site.

\begin{figure}[t]
\includegraphics[width=1\linewidth]{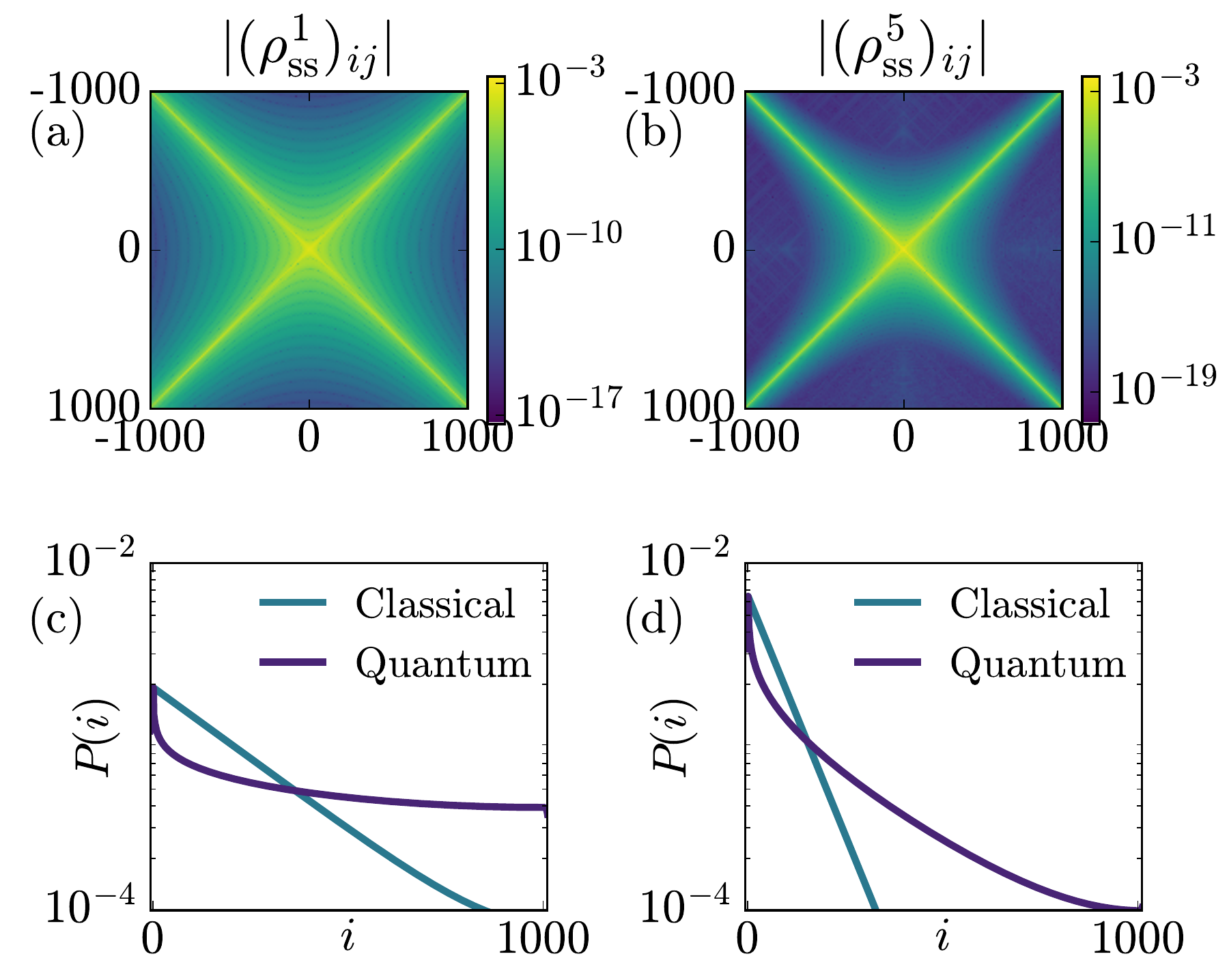}
\caption{(a,b) Magnitude of the stationary density matrix on a log scale. (c,d) Occupation probability for each site of the chain compared with the probability of a classical hopping model with exponential distribution. Plots are for (a,c) $\Gamma=1$ and (b,d) $\Gamma=5$.} 
\label{CoherentHoppingFig}
\end{figure}

For closed quantum systems, the spectrum of the corresponding Lindblad equation is given by all possible outer products of the eigenvectors and corresponding differences of eigenvalues of $H$. Given
\begin{equation}
H\ket{e_k}=\lambda_k\ket{e_k},
\end{equation} 
the matrices $R_{kk'}=\ket{e_k}\bra{e_{k'}}$ then provide the right eigenmodes of the Lindblad equation
\begin{equation}
\cL(R_{kk'})=-i(\lambda_{k}-\lambda_{k'})R_{kk'},
\end{equation} 
with the corresponding left eigenmodes also given by $L_{kk'}=\ket{e_k}\bra{e_{k'}}$. 
In this context, our results of Sec.~\ref{secresqtheory} can be modified to handle degenerate modes with $\cL(R_i)=0$, as is the case for closed quantum systems, analogously to the procedure described at the end of Sec. \ref{StatStateResets}. The resulting modification to Eq.~\eqref{SteadyStateResets} for coherent dynamics contains a sum over the non-zero eigenvalue modes as before, with the unique stationary density matrix replaced by a diagonal matrix in the energy eigenbasis of probabilities for the reset state to be measured in each eigenstate of the Hamiltonian. 

Using the outer product structure of the eigenstates, we can rewrite this sum more compactly as the matrix product 
\begin{equation}
\rho_{ss}^{\Gamma}=E\Lambda^\Gamma E^\dagger,
\end{equation} 
where $E$ is the matrix of eigenvectors defined by $HE=E\Lambda$ with $\Lambda_{ii}=\lambda_i$, and $\Lambda^\Gamma$ has elements defined by
\begin{equation}
\Lambda^\Gamma_{ij}=\frac{\Gamma\left\langle\psi|e_j\right\rangle\left\langle e_i|\psi\right\rangle}{\Gamma+i(\lambda_i-\lambda_j)}.
\end{equation}
These matrices can be efficiently constructed numerically and used to calculate the stationary state of a closed system after the addition of resets. 

For the coherent hopping model we have 
\begin{equation}
\ket{e_n}=\frac{1}{\sqrt{L}}\sum_{x=1}^{L}e^{i\frac{2\pi n}{L}(x-1)}\ket{x},
\end{equation}
and $\lambda_k=2\Omega \text{cos}(2\pi n/L)$. Choosing resets to the state $\ket{0}$ for a chain of $L=2001$ and hopping rate $\gamma=1000\gg\Gamma$, the numerically calculated stationary state is given for two different reset rates in Fig.~\ref{CoherentHoppingFig}. As expected, the magnitude of the components of the stationary state decay away from the reset state (Fig.~\ref{CoherentHoppingFig}\textcolor{blue}{(a,b)}). In Fig.~\ref{CoherentHoppingFig}\textcolor{blue}{(c,d)}, we plot the probability for the system to be found in each site against the distance from the reset state. For comparison, we also plot the probability for a classical random walk with resets, with parameters fixed by equating the probability of the two distributions at the reset state. We see that the coherent dynamics allows particles to move away from the reset state at a faster rate than the dissipative dynamics, leading to a crossing point beyond which there is a higher probability of locating the particle in the coherent model compared to the dissipative model. 

\subsection{Open quantum Ising model}

We consider as a second example the transverse field Ising model 
\begin{equation}
\label{Hamiltonian}
H = \Omega\sum_{j=1}^{N}{S}^{(j)}_{x}+V\sum_{i=1}^{N}{S}^{(j)}_{z}{S}^{(j+1)}_{z}
\end{equation}
with periodic boundary conditions, where the spin operators are ${S}^{(j)}_{\alpha}=\frac{1}{2}{\sigma}^{(j)}_{\alpha}$ with $\alpha=\{x,y,z\}$ and the jump operators are given by
\begin{equation}
\label{Jumps}
{J}_{j}=\sqrt{\kappa} \, {S}^{(j)}_{-}=\sqrt{\kappa} \, ({S}^{(j)}_{x}-i{S}^{(j)}_{y}).
\end{equation}
Unlike the models previously considered, this system is not exactly solvable without resets; however, it possesses a translation symmetry, which can be used with Eq.~\eqref{SteadyStateResets} to numerically diagonalise the model for larger system sizes than if we simply tried to diagonalize the symmetry-lacking Lindblad equation with resets.

An interesting feature of this model is the presence of metastability located in a region around a crossover of the stationary properties \cite{Rose2016}. This metastability takes the form of a decomposition of the system's state after a long evolution into a linear combination of a paramagnetic phase and ferromagnetic phase on either side of the crossover. This is followed by an eventual relaxation to a particular mixture of these two phases. We thus use this model to study the effect of adding resets to a model with metastability explicitly, comparing the quantum generalization of the results from Sec. \ref{Metastability} with the stationary states given by Eq.~\eqref{SteadyStateResets}.

\begin{figure}[t]
\includegraphics[width=1\linewidth]{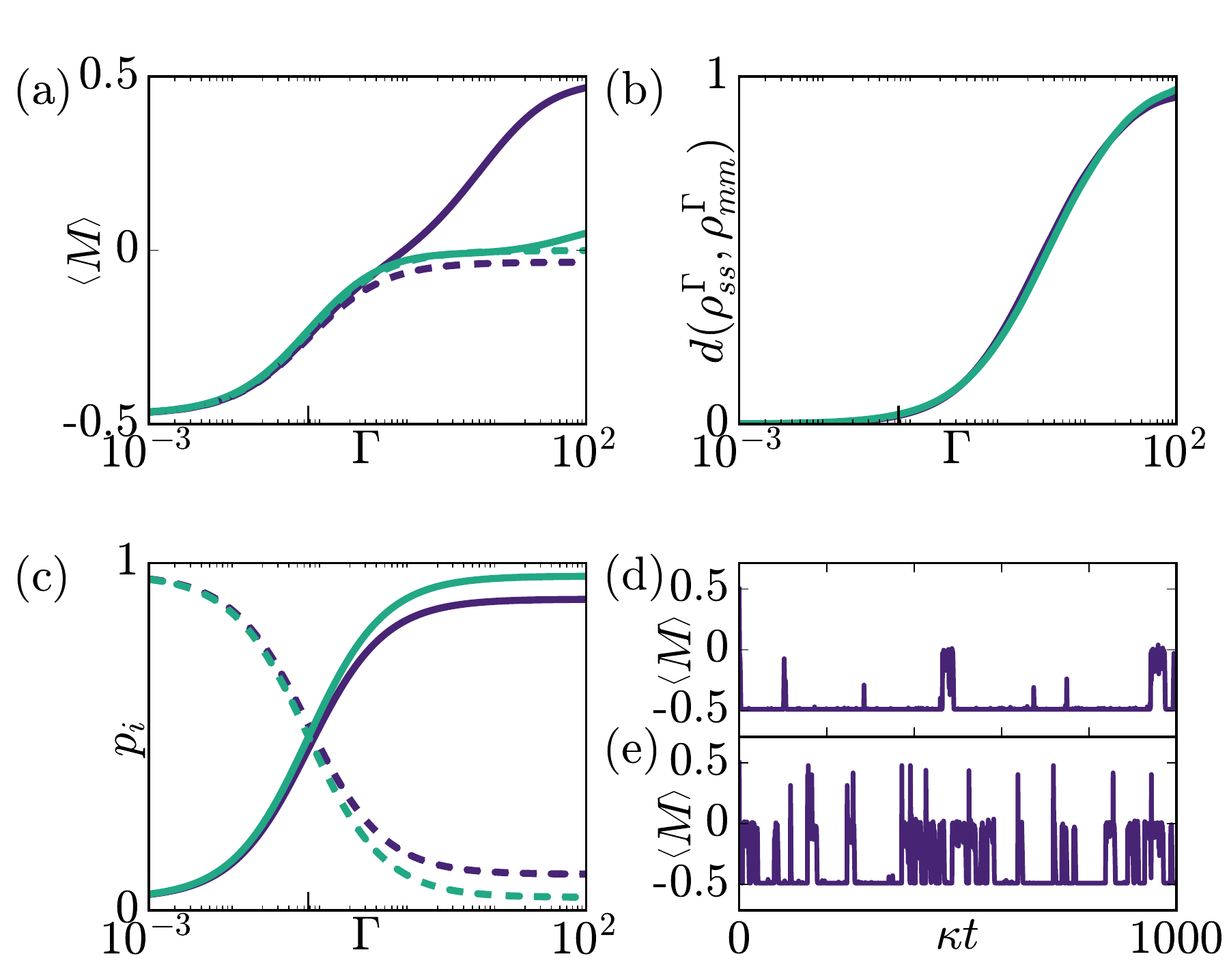}
\caption{(a) Magnetisation in the $z$ direction of the stationary state (solid) compared with the metastable approximation (dashed) as a function of the reset rate $\Gamma$. (b) Probability of the two extreme metastable states as a function of $\Gamma$. (c) Trace distance between the stationary state and the metastable approximation as a function of $\Gamma$. (d-e) Sample trajectories of the magnetisation over time for $\Gamma=0$ (d) and $\Gamma = 0.05$ (e). Purple lines are for resets to the ``all up'' state $\psi_1$, while blue lines are for resets to the ``alternating'' state $\psi_2$. Arrows indicate the point where $\Gamma = |\lambda_2|$.}
\label{OpenIsingFig}
\end{figure}

We start by studying the reset rate dependence of the system's magnetisation in the $z$ direction:
\begin{eqnarray}
M=\frac{1}{N}\sum_{i=1}^{N}S_z^i.
\end{eqnarray}
Considering a system of $N=7$ spins, in Fig.~\ref{OpenIsingFig}\textcolor{blue}{(a)} we plot the magnetisation of both the full reset stationary state given by Eq.~\eqref{SteadyStateResets} (solid lines) and the approximate decomposition of the reset stationary state into the original metastable phases given by Eq.~\eqref{ResetMetastableDecomp} (dashed lines). This is done for two different reset states, $\ket{\psi_1}=\ket{\uparrow\uparrow\uparrow\uparrow\uparrow\uparrow\uparrow}$ and $\ket{\psi_2}=\ket{\uparrow\downarrow\uparrow\downarrow\uparrow\downarrow\uparrow}$, both of which have a high probability of evolving into the paramagnetic state after a time in the metastable regime.

Without resets the stationary state is dominated by the ferromagnetic phase. However, for both reset states, there is a larger probability of evolving into the paramagnetic phase than the ferromagnetic phase on metastable timescales. When the reset rate is increased, we expect the stationary state to become more biased towards the paramagnetic phase. This is clearly demonstrated in Fig.~\ref{OpenIsingFig}\textcolor{blue}{(a)}, with strong agreement between the exact result and the approximation for reset rates up to the order of $|\lambda_2|$. For $\Gamma$ beyond this scale, metastability is lost and the approximation fails, with the magnetisation approaching that of the reset states for large $\Gamma$. 

To quantify the agreement, we show in Fig.~\ref{OpenIsingFig}\textcolor{blue}{(b)} the trace distance between the truncated metastable state, as given by Eq.~\eqref{ResetMetastableDecomp}, and the full reset stationary state, as given by Eq.~\eqref{SteadyStateResets}. We can see that this distance is close to zero up until $\Gamma=|\lambda_2|$, after which it increases rapidly, demonstrating a strong accuracy of Eq.~\eqref{ResetMetastableDecomp} when the average time between resets is equal to or longer than the metastable timescale, as assumed in Sec.~\ref{Metastability}. This change in the stationary magnetisation for smaller reset rates corresponds directly to the changing mixture of metastable phases in the stationary state. This is seen in Fig.~\ref{OpenIsingFig}\textcolor{blue}{(c)}, which shows a higher probability of the system being found in the paramagnetic phase as $\Gamma$ is increased. The same behavior can also be seen at the trajectory level in the two plots of Fig.~\ref{OpenIsingFig}\textcolor{blue}{(d,e)}, which show sample trajectories without resets and with resets to the state $\ket{\uparrow\uparrow\uparrow\uparrow\uparrow\uparrow\uparrow}$ at a rate $\Gamma$, respectively. As expected, we see that resets induce more periods of paramagnetic phase dropping back into the ferromagnetic phase.

We note that it may be possible to conduct an experiment to study the effect of resets on this Ising model (and similar many-body spin models) using ultra-cold atoms confined in optical latices \cite{Muller2008,Saffman2010,Viteau2011,Lee2012}. In such experiments, the $\ket{\downarrow}$-state is associated with the atomic ground state, while the $\ket{\uparrow}$-state is represented by a Rydberg nS-state. These states are coupled coherently by laser, leading to Rabi oscillations at a frequency $\Omega$ and a detuning $\Delta$ relative to the energy difference between the two states. Excited atoms at lattice sites with position $r_i$ and $r_j$ interact via a van der Waals potential $V_{ij}=C_6/|r_i-r_j|^6$, where $C_6$ is the dispersion coefficient characterizing the interaction strength. Altogether this gives a Hamiltonian of the form
\begin{equation}
H = \Omega\sum_iS_x^i+\Delta\sum_iS_z^i+\frac{1}{2}\sum_{i\neq j}V_{ij}n_in_j,
\end{equation}
where $n_i=1/2I+S^i_z$. For sufficiently large lattice spacing $a$, the interaction decays so rapidly that it can be approximated as a nearest-neighbour interaction. Applying a laser detuning of the form $\Delta=-C_6/a^6$ then leads for a periodic chain to the Ising Hamiltonian \eqref{Hamiltonian}, up to an overall energy shift that can be discarded. Dissipation occurs naturally via photon-emitted decay of the Rydberg states.

To simulate resets in this system, we can force it into a high magnetisation state at random intervals determined externally from an exponential distribution. While it is difficult in practice to place the system in a specific pure state with high magnetisation reliably, the above results simply generalize to a probability distribution of pure reset states (i.e., a reset ``density matrix''). If such a density matrix has a large positive expectation value for the magnetisation, evolution after reset will have a high probability of leading to the paramagnetic state on the metastable timescale, resulting in similar observations to those of Fig.~\ref{OpenIsingFig}. These resets could be implemented via a strong laser pulse, such that the system can be momentarily approximated as non-interacting and the pulse modelled as instantaneous.

\section{Conclusion}

We have developed in this work a general spectral approach to investigate the properties of Markov processes that are reset to a fixed state at random exponentially distributed times. Our main result shows that the spectrum of the generator of a reset Markov process is globally shifted by the reset rate compared with the spectrum of the corresponding reset-free process, except for the stationary mode, which stays at zero. We have also provided an explicit formula for the stationary distribution of the reset process, based on the spectrum of the reset-free process. 

This spectral approach can be applied not only to classical stochastic processes but also, as we have shown, to closed and open quantum systems modelled by Lindblad-type equations. In both cases, the approach provides a natural way to study how resets can create a stationary state by opening a gap in the spectrum and how it affects metastable states. This was illustrated using various classical and quantum processes, including reset Brownian motion and the transverse field Ising model reset to a paramagnetic or ferromagnetic state.

For future work, it would be interesting to develop a similar approach for the large deviations of reset processes, based on the spectrum of the tilted generator \cite{meylahn2015b,Harris2017}. There does not seem to be, a priori, a direct extension of our results for this generator, as the basic property used to prove our results, namely, that the non-stationary eigenstates have zero norm, does not hold in general for the eigenstates of the tilted generator. However, it might be possible to obtain partial results when the addition of reset to the master operator mixes only a small subset of the non-reset spectrum. 

There also remains much work to be done on quantum systems, for which resets can be induced either by measurements, as in \cite{{Thiel2017,Friedman2017}}, or through other external perturbations. Some results on these systems have also been obtained very recently in \cite{mukherjee2018} for the stationary state of closed quantum systems with reset, which form, as we have seen, a specific case of the quantum systems considered here.

\begin{acknowledgments}
This work was supported by ERC under the EU Seventh Framework Programme (Grant No.\ FP/2007-2013), the ERC project ESCQUMA (Grant No.\ 335266), and by EPSRC (Grants No.\ EP/M014266/1 and No.\ EP/R04340X/1). H.T.\ is supported by NRF South Africa (Grant No.\ 96199) and by Stellenbosch University. I.L. gratefully acknowledges funding through the Royal Society Wolfson 
Research Merit Award. This work was also supported in part by the International Centre for Theoretical Sciences (ICTS) during a visit for participating in the program ``Large deviation theory in statistical physics: Recent advances and future challenges'' (Code: ICTS/Prog-ldt/2017/8).
\end{acknowledgments}
	
%
\end{document}